\documentclass[%
 reprint,
 amsmath,amssymb,
 aps,
]{revtex4-1}

\usepackage{graphicx}
\usepackage{dcolumn}
\usepackage{bm}
\usepackage{subfigure}

\begin{document}


\title{Tunable photon blockade in a whispering-gallery-mode microresonator coupled with two nanoparticles}

\author{Wen-An Li}
 \email{liwenan@126.com, liwa@gzhu.edu.cn}
\affiliation{School of Physics and Electronic Engineering, Guangzhou University, Guangzhou 510006, China}


\date{\today}

\begin{abstract}
We have theoretically studied the photon statistical properties in a nonlinear whispering-gallery-mode microresonator coupled with two nanoparticles. By tuning the relative position of two nanoparticles, the photon statistical features of the system can be modified remarkably. Interestingly, a controllable switching between unconventional and conventional photon blockade can be realized by manipulating the angular positions of two nanoparticles. We also investigate the influence of the Kerr effect on the second order correlation function and find that there is an optimal choice for the relative position of two nanoparticles and the strength of Kerr effect that can generate strong antibunching. Furthermore, under the strong driving, two photon blockade can be achieved when the system is close to an exceptional point. Our work may provide an effective way to control photon statistical characteristics and have potential applications in quantum information science.
\end{abstract}

\pacs{Valid PACS appear here}
\maketitle

\section{introduction}
A single photon source, an indispensable device for generating photons at the single-photon level, plays a central role in diverse areas such as quantum cryptography~\cite{1}, quantum information processing~\cite{2}, single-photon transistor~\cite{3}, quantum computation~\cite{4} and so on. The photon blockade is one of most attractive mechanisms for constructing a single photon source. In close analog to Coulomb blockade for electrons~\cite{5,6}, the photon blockade is a striking quantum phenomenon, where the excitation of a photon blocks the transport of subsequent photons for the nonlinear cavity so that they are emitted one by one. As a consequence, the cavity can only host one photon at a time, acting as a ``photon turnstile''~\cite{7,8}. In 2005, the photon blockade was first demonstrated experimentally with a single atom trapped in an optical cavity~\cite{9}. Subsequently, the strong antibunching behaviors were predicted in various experimental setups including a quantum dot in a photonic crystal~\cite{10,11} and circuit cavity quantum electrodynamics systems~\cite{12,13,14}. In these works, the observation of antibunching requires large nonlinearities with respect to the decay rate of the system, so it is known as ``conventional photon blockade''.

Apart from the single photon blockade, the multi-photon blockade has also attracted much interest due to its potential applications in multiphoton quantum-nonlinear optics like an $n$-photon source ($n>1$)~\cite{15}. To date, the multi-photon blockade has been studied in various configurations~\cite{16,17,18,19,20}. For instance, the two- and three-photon blockade can be observed in a system consisting of a cavity with Kerr nonlinearity driven by a weak classical field~\cite{16}. The prerequisite for realizing multi-photon blockade in this system is the presence of strong nonlinearities. Another method to realize three-photon blockade is based on the collective decay of two atoms trapped in a single-mode cavity with different coupling strengths~\cite{17}. In this scheme, the two-photon and three-photon blockades strongly depend on the location of two atoms in the strong-coupling regime. Recently, the two-photon blockade was first observed in an atom-driven cavity quantum electrodynamics system~\cite{21}. Although many progress on the study of multi-photon blockade has been made, the accomplishment of the multi-photon blockade is still challenging in experiments.   

In 2010, Liew and Savona found a new mechanism for the photon blockade, where strong photon antibunching can be obtained even with nonlinearities much smaller than the decay rates of the cavity modes~\cite{22}. This mechanism is referred to as the ``unconventional photon blockade''. Its feature can be understood as destructive quantum interference between different excitation pathways from the ground state to the two-photon states. Since then, a sequence of theoretical proposals based on this mechanism were suggested in many different systems including, for example, a bimodal optical cavity with a quantum dot~\cite{23,24,25,26,27,28,29}, symmetric and antisymmetric modes in weakly nonlinear photonic molecules~\cite{30}, coupled optomechanical systems~\cite{31,32}. More recently, the unconventional photon blockade was experimentally observed in two coupled superconducting circuit resonators~\cite{33} and in a quantum dot embedded in a bimodal micropillar cavity~\cite{34}.

In parallel, the physical systems described by non-Hermitian Hamiltonians have also attracted much interest~\cite{rev1,rev2,rev3,rev4,rev5}, because such Hamiltonians exhibit special degeneracies known as exceptional points, at which two or more eigenvalues and the corresponding eigenvectors coalesce. In 2001, the physical existence of the exceptional point was experimentally demonstrated in microwave cavities~\cite{35}. Subsequently, a variety of unconventional effects have been observed in experiments, such as loss-induced coherence~\cite{36,37}, unidirectional lasing~\cite{38}, wireless power transfer~\cite{39}, and exotic topological states~\cite{40,41}. In recent experiments~\cite{42,43,44}, by coupling two nanoscale scatters (i.e. nanoparticles) to a whispering-gallery-mode (WGM) micro-toroid cavity, the system can be steered in a precise and controlled manner to the exceptional point.  The presence of two nanoparticles within the mode volume of the cavity leads to the asymmetric backscattering of counter-propagating optical waves, which can be adjusted by manipulating the relative position of two nanoparticles. In the vicinity of the exceptional points, some counterintuitive effects have been shown including loss-induced revival of lasing~\cite{37}, ultra-sensitive sensor~\cite{42}, chiral lasing~\cite{43} and optomechanically induced transparency~\cite{46}. 

Motivated by above works~\cite{42,43,44,46}, one question that arises naturally is whether the asymmetric coupling of two counter-propagating optical waves affects the photon statistical properties of cavity modes. In the previous works~\cite{25,26,27,30}, studies on the unconventional photon blockade are based on the \emph{symmetric} coupling of the optical modes. According to the optimal conditions, the required Kerr nonlinearity decreases with increasing coupling strength of the optical modes. It means that strong photon antibunching with weak Kerr nonlinearity requires large optical coupling between optical modes, which is not easy to realize in the experiments. Here, we consider the \emph{asymmetric} coupling of two optical modes in one resonator and study the new possibility of controlling the photon blockade by tuning the relative angular position of two nanoparticles along the circumference of the nonlinear microresonator.  In fact, adjusting the relative position of two nanoparticles corresponds to the change in the relative phase of the coupling coefficients without increasing the amplitudes of  coupling constants. We find that the relative phase of the coupling coefficients plays a crucial role in modifying the photon statistical properties of the system. By tuning the relative position of two nanoparticles, the photon statistical properties can be well controlled and the switching between unconventional and conventional photon blockade can be realized. We also investigate the influence of the Kerr nonlinearity strength on the photon statistics properties. Furthermore, in the vicinity of an exceptional point, two-photon blockade effect can be achieved under the strong driving. Our work, with weak nonlinearity but without requiring strong coupling between optical modes, can be realized within current experimental techniques.

The remainder of the paper is organized as follows. In Sec. II, the theoretical model and Hamiltonian are described for the nonlinear WGM microresonator coupled with two nanoparticles. In Sec. III, the output power spectra of the WGM microresonator system are presented. Subsequently, in Sec. IV, the photon statistical properties of present system is analytically and numerically discussed. Finally, a summary of the main results is given in Sec. V.
\begin{figure}[t]
\begin{center}
\includegraphics[width=0.4\textwidth]{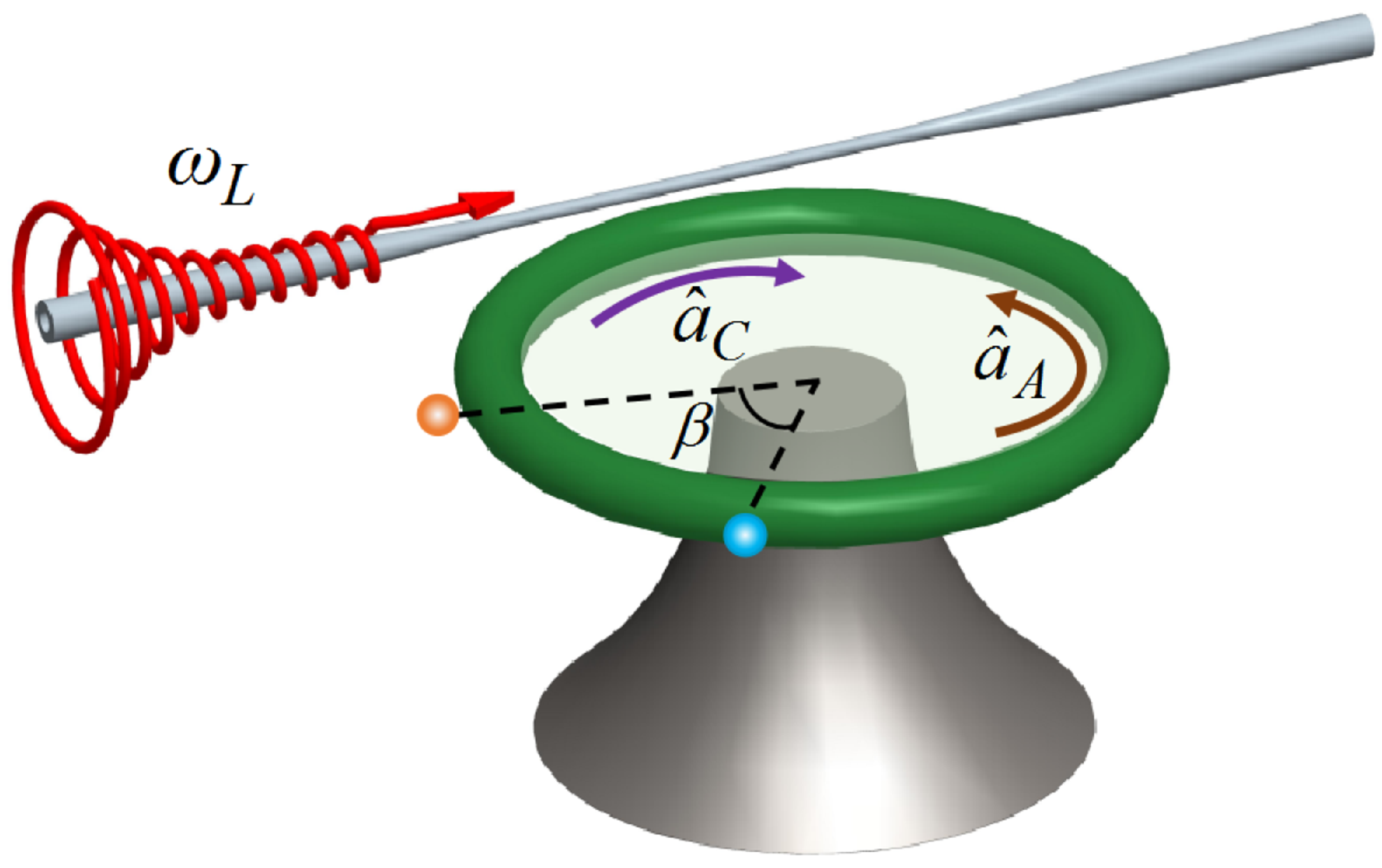}
\end{center}
\caption{Schematic diagram of the WGM microresonator with Kerr medium coupled with two nanoparticles which is coherently driven by a pump field at frequency $\omega_L$ through an optical tapered fiber waveguide. The WGM microresonator supports two counterpropagating modes (clockwise mode $\hat{a}_{\mathrm{C}}$ and anti-clockwise mode $\hat{a}_{\mathrm{A}}$), which can be asymmetrically coupled through backscattering by two nanoparticles. $\beta$ is the relative angle between two particles.}
\label{fig1}
\end{figure}

\section{theoretical model}
As shown in Fig.\ref{fig1}, we consider a WGM microresonator with Kerr medium coupled to an optical fiber waveguide for in- and out-coupling of light. With its circular geometry, the WGM cavity supports clockwise and anti-clockwise travelling modes with degenerate eigenfrequencies $\omega_c$ and the same decay rate $\gamma=\gamma_{\mathrm{ex}}+\gamma_{\mathrm{in}}$. $\gamma_{\mathrm{ex}}$ is the external decay rate (the outgoing coupling coefficient) from the WGM microresonator into the tapered fiber and $\gamma_{\mathrm{in}}$ is the intrinsic decay rate. Two nanoparticles are placed in the evanescent field of the resonator, which can tune the coherent backscattering of clockwise and anti-clockwise travelling modes inside the resonator. In the presence of the optical loss, the system considered here is an open system and the Hamiltonian is non-Hermitian. 
\subsection{\label{sec:level2}Review of the two-mode approximation model}
In order to give a full description of this open system, we first briefly review the two-mode approximation model and the eigenmode evolution in a WGM microresonator with nanoscatter-induced broken spatial symmetry. The two-mode approximation model was first phenomenologically introduced for deformed microdisk cavities~\cite{tma1,tma2} and was later rigorously derived for the microdisk with two scatterers~\cite{47}. The key idea is to model the dynamics in the slowly-varying envelope approximation in the time domain with a Schr\"{o}dinger-like equation $id\Psi/dt=H\Psi$. Here, $\Psi$ is the two-component column vector $(\Psi_A, \Psi_C)^T$, where the superscript $T$ indicates the matrix transpose. The complex-valued entry $\Psi_A$ ($\Psi_C$) stands for all the field amplitudes of the anti-clockwise (clockwise) propagating waves. Since the microcavity is an open system, the corresponding effective Hamiltonian,
\begin{equation}
H=
\left(
\begin{matrix}
 \Omega     & J_1        \\
J_2      &  \Omega         
\end{matrix}
\right)
\end{equation}
is a $2\times2$ matrix, which is in general non-Hermitian.
The real parts of the diagonal elements $\Omega$ are the frequencies and the imaginary parts are the decay rates of the resonant traveling waves. The complex-valued off-diagonal elements $J_1$ and $J_2$ are the backscattering coefficients, which describe the scattering from the clockwise (anti-clockwise) to the anti-clockwise  (clockwise) travelling wave. In general, in the open system the backscattering is asymmetric, $|J_1|\neq |J_2|$, which is allowed because of the non-Hermiticity of the Hamiltonian. A short calculation shows that the complex eigenvalues of $H$ are $\Omega_{\pm}=\Omega\pm\sqrt{J_1J_2}$ and the complex (not normalized) right eigenvectors are
\begin{equation}
\Psi_{\pm}=
\left(
\begin{matrix}
 \sqrt{J_1}     \\
 \pm\sqrt{J_2}              
\end{matrix}
\right).
\end{equation}
Clearly, in the case of asymmetric backscattering one component of a given eigenvector is larger than the other component. Physically, it means that the eigenvectors show an imbalance of clockwise and anti-clockwise components if the backscattering is asymmetric. For the particular case of the WGM microresonator perturbed by two scatterers the matrix elements of $H$ are determined as follows~\cite{42,43,47},
\begin{subequations}
\begin{align}
&\Omega=\omega_c-i\frac{\gamma}{2}+\sum_{j=1}^2\epsilon_j, \\
&J_1=\sum_{j=1}^2\epsilon_j e^{-i2m\beta_j}, \\
&J_2=\sum_{j=1}^2\epsilon_j e^{i2m\beta_j},
\end{align}
\end{subequations}
where $m$ is the azimuthal mode number, $\beta_j$ is the angular position of scatterer $j$ and $2\epsilon_j$ is the complex frequency splitting that is introduced by scatterer $j$ alone. $\epsilon_j$ can be calculated for the single-particle-microdisk system either fully numerically (using, e.g., the finite element method (FEM)~\cite{fem}, the boundary element method (BEM)~\cite{bem}), or analytically using the Green's function approach~\cite{green}. In recent experiments, $\epsilon_j$ can be adjusted by tuning the distance between the resonator and the particles. Here, we take the position of one of the nanoparticles as the reference position. For example, take the orange particle (in Fig.1) as the first particle and set its angular position to be $\beta_1=0$, then the angular position of the second particle is $\beta_2=\beta$, where $\beta$ represents the relative angular position of the two scatters. Therefore, the asymmetric backscattering coefficients of counterpropagating waves, induced by the nanoparticles, can be reduced to 
\begin{equation}
J_{1,2}=\epsilon_1+\epsilon_2 e^{\mp i 2m\beta}.
\end{equation}
It is noted that the relative angular $\beta$ is of great importance, since it can modify the photon statistical properties of the system (see discussions below). 

Although the two-mode approximation model was given for the isolated microdisk cavity perturbed by two particles, it is still valid in the waveguide-cavity systems by assuming that there is no backscattering of light between the microcavity and the waveguides. It can be justified when the distances between cavity and waveguides are sufficiently large. Note that, the extension of the two-mode model to waveguide-cavity systems has been introduced and tested in recent experiments~\cite{42,43}.
\subsection{\label{sec:level2}The Hamiltonian of our model}
Based on above discussions, we will give a theoretical description of our model. To make our scheme work, a driving laser of frequency $\omega_L$ is applied to the system via the evanescent coupling of the optical fiber and the resonator, the field amplitudes are given by $F=\sqrt{\gamma_{\mathrm{ex}} P_L/\hbar \omega_L}$, where $P_L$ is the pump power. In the frame rotating with the input field frequency $\omega_L$, the Hamiltonian of this system is described by
\begin{align}\label{eq1}
\nonumber
\hat{H}_{\mathrm{sys}}=&\Delta(\hat{a}_{\mathrm{C}}^\dag \hat{a}_{\mathrm{C}}+
 \hat{a}_{\mathrm{A}}^\dag \hat{a}_{\mathrm{A}})+U (\hat{a}_{\mathrm{C}}^\dag \hat{a}_{\mathrm{C}}^\dag \hat{a}_{\mathrm{C}}\hat{a}_{\mathrm{C}}\\ \nonumber
&+\hat{a}_{\mathrm{A}}^\dag \hat{a}_{\mathrm{A}}^\dag \hat{a}_{\mathrm{A}}\hat{a}_{\mathrm{A}})
+J_1\hat{a}_{\mathrm{C}} \hat{a}_{\mathrm{A}}^\dag+J_2\hat{a}_{\mathrm{C}}^\dag \hat{a}_{\mathrm{A}}  \\ 
&+iF(\hat{a}_{\mathrm{C}}^\dag-\hat{a}_{\mathrm{C}})
\end{align}
where $\Delta=\Delta_c+\mathrm{Re}(\epsilon_1+\epsilon_2)$, and $\Delta_c=\omega_c-\omega_L$. The nonlinear Kerr coefficient is given by $U=\hbar \omega_c^2 c n_2/n_0^2V_{\mathrm{eff}}$, where $c$ is the speed of light in vacuum, $n_0$ and $n_2$ are the linear and nonlinear refractive index of the material and $V_\mathrm{eff}$  is the effective mode volume. $\hat{a}_{\mathrm{C}}$ ($\hat{a}_{\mathrm{A}}$) and $\hat{a}_{\mathrm{C}}^\dag$ ($\hat{a}_{\mathrm{A}}^\dag$) are the photon annihilation and creation operators of the clockwise modes (anti-clockwise modes), satisfying the bosonic commutation relations $[\hat{a}_{\mathrm{C}},\hat{a}_{\mathrm{C}}^\dag]=1$ and $[\hat{a}_{\mathrm{A}},\hat{a}_{\mathrm{A}}^\dag]=1$.

In above Hamiltonian (\ref{eq1}), the first term denotes the energy of the WGM microresonator in the rotating frame. The second term represents the Kerr nonlinear interaction. The third and fourth terms are the coherent coupling of the clockwise mode with anti-clockwise mode.  In general, $J_1\neq J_2$, which can be tuned by the relative angular position of two nanoparticles and the distance between nanoparticles and the WGM microresonator. Due to this asymmetric coupling between two counterpropagating modes, some interesting, controllable photon statistical properties will be shown in our system. The last term describes the interaction between the cavity field and the input field.

\begin{figure}
\begin{center}
\subfigure[]{\includegraphics[width=0.4\textwidth]{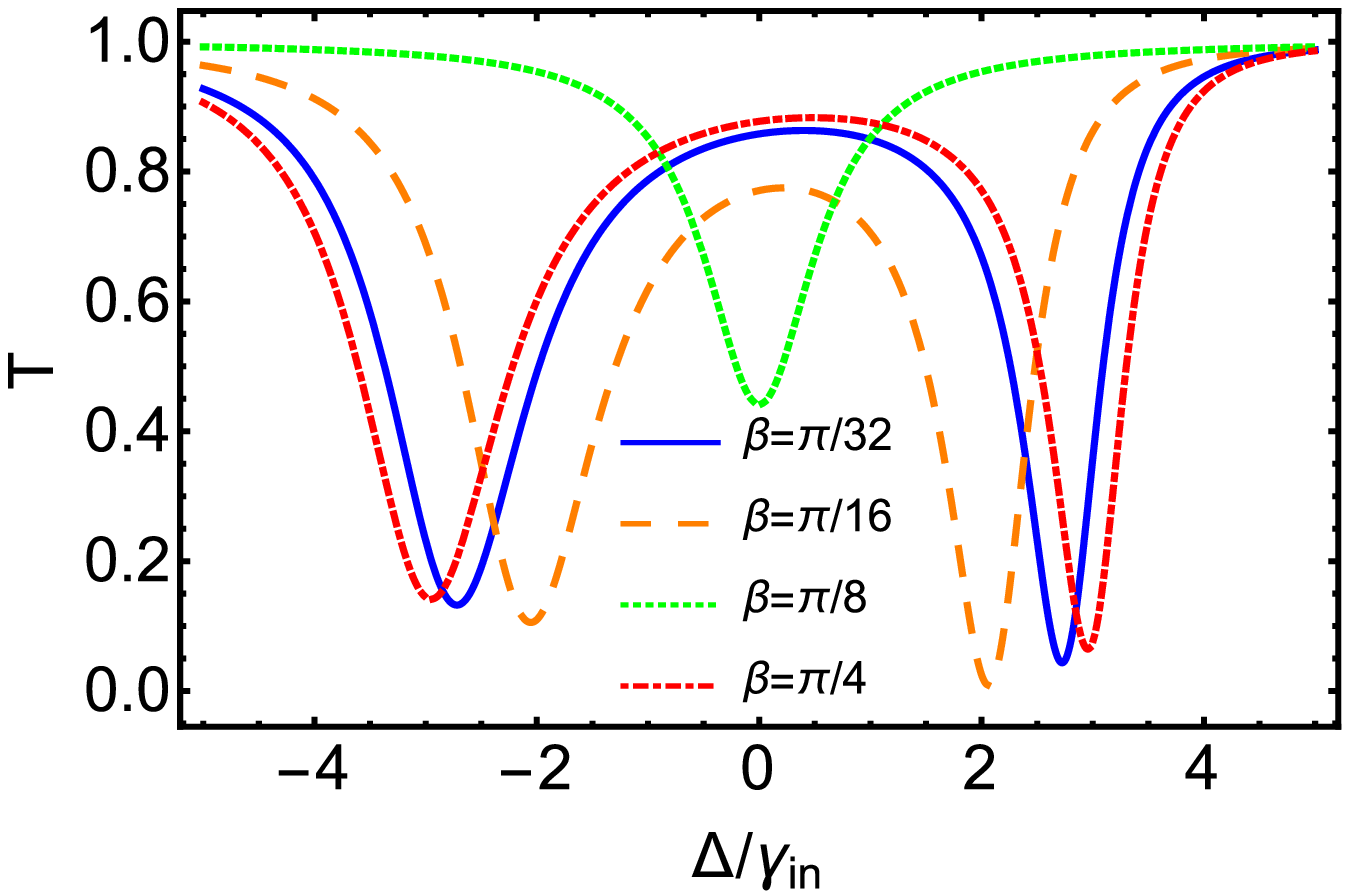}}
\subfigure[]{\includegraphics[width=0.42\textwidth]{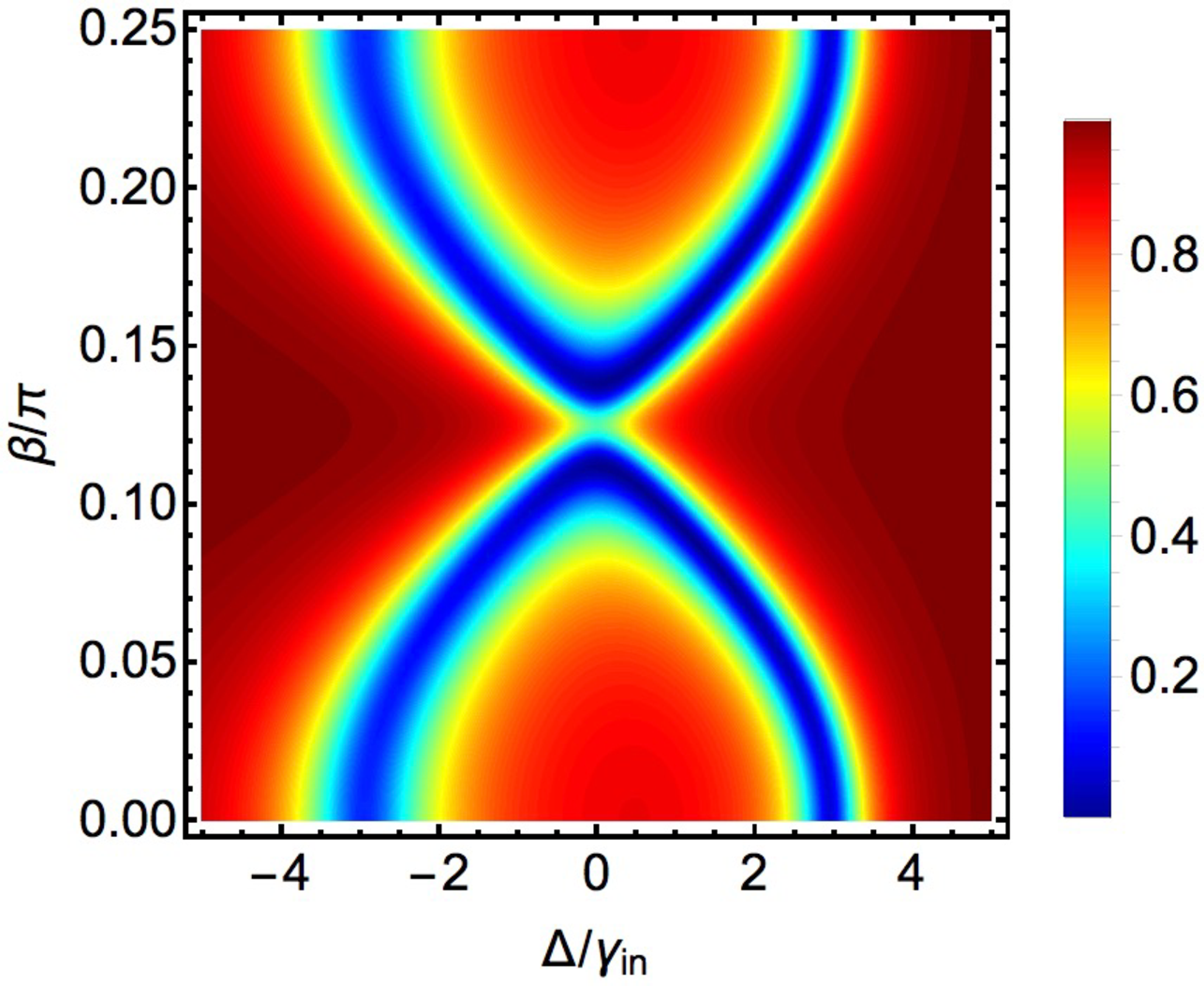}}
\end{center}
\caption{(a) The system output transmission spectra as a function of the optical detuning $\Delta/\gamma_{\mathrm{in}}$ under different relative angle $\beta$. (b) Transmission as a function of the optical detuning 􏰟$\Delta/\gamma_{\mathrm{in}}$ and the angle $\beta$. Here, we have selected $\gamma_{\mathrm{ex}}/\gamma_{\mathrm{in}}=1$, $\epsilon_1/\gamma_{\mathrm{in}}=1.5-0.1i$, $\epsilon_2/\gamma_{\mathrm{in}}=1.4999-0.101489i$, $U/\gamma_{\mathrm{in}}=0.059$, $F/\gamma_{\mathrm{in}}=0.01$ and $m=4$.}
\label{fig2}
\end{figure}
\section{output power spectra of the WGM microresonator}
Before discussing the photon statistical properties of our system, we first study the system output power spectra. As mentioned above, when two nanoparticles are placed along the circumference of the resonator, the system exhibits fully asymmetric internal backscattering. The position of each particle can be controlled by a nanopositioner, which tunes the relative position and effective size of the nanotip in the WGM fields~\cite{42,43}. By carefully tuning the relative positions of two particles, the system can be steered to an exceptional point. For the present system, the nanoparticles induced frequency splitting of the optical modes can be derived as $\Delta\omega=\pm\sqrt{J_1J_2}$, thus the corresponding critical value of $\beta$ can be obtained as~\cite{46}
\begin{equation}
\beta_c=\frac{l\pi}{2m}\mp\frac{\mathrm{arg}(\epsilon_1)-\mathrm{arg}(\epsilon_2)}{2m} \quad (l=\pm1,\pm3,...),
\end{equation}
where $\mp$ corresponds to $J_1=0$ or $J_2=0$. Here, $\epsilon_j$ ($j=1,2$) are complex numbers, and we assume $|\epsilon_1|=|\epsilon_2|$. $\mathrm{arg}(\epsilon_j)$ denotes the argument of complex number $\epsilon_j$.
In the vicinity of the exceptional points, some unconventional effects may occur. Thus, it is of great interest to investigate the output power spectra of such coupled system when the relative position of two particles varies.

According to the Hamiltonian (\ref{eq1}) above, the dynamics of the coupled system can be described by the quantum Langevin equations
\begin{subequations}\label{eq3}
\begin{align}
\frac{d}{dt}\hat{a}_{\mathrm{C}}=&\left(-\frac{\gamma_{\mathrm{opt}}}{2}-i\Delta-2iU\hat{a}_{\mathrm{C}}^\dag \hat{a}_{\mathrm{C}}\right)\hat{a}_{\mathrm{C}}-iJ_2\hat{a}_{\mathrm{A}}+F+\hat{a}_{\mathrm{in}}^{\mathrm{C}},\\
\frac{d}{dt}\hat{a}_{\mathrm{A}}=&\left(-\frac{\gamma_{\mathrm{opt}}}{2}-i\Delta-2iU\hat{a}_{\mathrm{A}}^\dag \hat{a}_{\mathrm{A}}\right)\hat{a}_{\mathrm{A}}-iJ_1\hat{a}_{\mathrm{C}}+\hat{a}_{\mathrm{in}}^{\mathrm{A}},
\end{align}
\end{subequations}
where $\hat{a}_{\mathrm{in}}^{\mathrm{C}}$ and $\hat{a}_{\mathrm{in}}^{\mathrm{A}}$ are the input vacuum noises of the cavity modes, respectively. $\gamma_{\mathrm{opt}}=\gamma_{\mathrm{in}}-\mathrm{Im}(\epsilon_1+\epsilon_2)$ is the total optical loss. Under the mean-field approximation, we assume that the mean values of these noise operators are zero, i.e., $\langle \hat{a}_{\mathrm{in}}^{\mathrm{C}}\rangle=\langle \hat{a}_{\mathrm{in}}^{\mathrm{A}}\rangle=0$. Here, we are interested in the influence of relative position of two nanoparticles on the system output power spectra. Under weak Kerr nonlinearity $U\ll \gamma_{\mathrm{in}}$, we can easily omit the Kerr interaction terms in above Eq.(\ref{eq3}). Furthermore, we assume that all of the time derivatives in the quantum Langevin equations are set to be zero. Thus, it is easy to obtain steady-state values of the dynamical variables as
\begin{subequations}
\begin{align}
\langle \hat{a}_{\mathrm{C}}\rangle=\frac{F(\gamma_{\mathrm{opt}}/2+i\Delta)}{(\gamma_{\mathrm{opt}}/2+i\Delta)^2+J_1J_2},\\
\langle \hat{a}_{\mathrm{A}}\rangle=\frac{-iFJ_1}{(\gamma_{\mathrm{opt}}/2+i\Delta)^2+J_1J_2},
\end{align}
\end{subequations}
Note that $J_1J_2=\epsilon_1^2+\epsilon_2^2+2\epsilon_1\epsilon_2\cos(2m\beta)$.
We can see that the $\beta$-dependent optical coupling rate indeed affects the intracavity optical intensity. By using the standard input-output relations, i.e., $\langle \hat{a}_{\mathrm{out}}\rangle=\langle \hat{a}_{\mathrm{in}}\rangle-\sqrt{\gamma_{\mathrm{ex}}}\langle \hat{a}_{\mathrm{C}}\rangle$, we obtain the the normalized power forward transmission spectra
\begin{equation}
T=\left|\frac{\langle \hat{a}_{\mathrm{out}}\rangle}{\langle \hat{a}_{\mathrm{in}}\rangle}\right|^2=\left|1-\frac{\gamma_{\mathrm{ex}}}{F}\langle \hat{a}_{\mathrm{C}}\rangle\right|^2.
\end{equation}
When the Kerr terms are included, the exact expression of $\langle \hat{a}_{\mathrm{C}}\rangle$ ($\langle \hat{a}_{\mathrm{A}}\rangle$)  can not be obtained generally. Therefore, we numerically calculate the solutions to equations (\ref{eq3}) under the mean-field approximation and plot the transmission rate versus detuning under different relative angular position of two particles in Fig.~\ref{fig2}(a). Here, we have selected the experimentally accessible values $\gamma_{\mathrm{ex}}/\gamma_{\mathrm{in}}=1$, $\epsilon_1/\gamma_{\mathrm{in}}=1.5-0.1i$, $\epsilon_2/\gamma_{\mathrm{in}}=1.4999-0.101489i$, $U/\gamma_{\mathrm{in}}=0.059$ and $m=4$~\cite{42,43,46}. With these parameters, the exceptional point corresponds to the angular position at $\beta_c\approx0.4$. Fig.~\ref{fig2}(a) shows the $\beta$-dependent transmission rate with two nanoparticles, featuring a asymmetric spectrum around the resonance due to the asymmetric backscattering between the clockwise- and anticlockwise-travelling waves. More interestingly, when $\beta$ is set to be $\pi/8$ (in the vicinity of the exceptional points), the transmission spectra demonstrates only one local minimum at the resonance. For $\beta=\pi/16$, strong absorption is shown around $\Delta/\gamma_{\mathrm{in}}=2$. However, by tuning the system close to the exceptional point (namely, $\beta=\pi/8$), a transparency window emerges. Hence, an optical switching (at $\Delta/\gamma_{\mathrm{in}}=2$) can be achieved by adjusting the relative angular position of two particles. For completeness, we plot the transmission spectra versus driving field detuning and relative angle between two nanoparticles in Fig.~\ref{fig2}(b).

\section{photon statistical properties of the WGM microresonator with two nanoparticles}
\subsection{\label{sec:level2}General formalism}
To correctly account for the driven-dissipative character of the system, we introduce the quantum master equation for the system density matrix,
\begin{align}\label{eq7}
\frac{d\hat{\rho}}{dt}=-i[\hat{H}_{\mathrm{sys}}, \hat{\rho}]+\gamma_1\mathcal{L}[\hat{a}_{\mathrm{C}}]\hat{\rho}+\gamma_2\mathcal{L}[\hat{a}_{\mathrm{A}}]\hat{\rho}
\end{align}
where $\mathcal{L}[\hat{x}]\hat{\rho}=\hat{x}\hat{\rho} \hat{x}^\dag-\frac{1}{2}\hat{x}^\dag \hat{x}\hat{\rho}-\frac{1}{2}\hat{\rho} \hat{x}^\dag \hat{x}$ is the Lindblad superoperator term for the collapse operator $\hat{x}$ acting on the density matrix $\hat{\rho}$ to account for losses to the environment. $\gamma_1$ and $\gamma_2$ denote the damping constant of clockwise mode and anti-clockwise mode, respectively. Here, the decay rates of the resonator modes are assumed to be equal, i.e., $\gamma_1=\gamma_2=\gamma_{\mathrm{opt}}$. The steady-state solution $\rho_{ss}$ of the density matrix $\hat{\rho}$ can be obtained by setting $d\hat{\rho}/dt=0$ in Eq. (\ref{eq7}). 
To observe the photon blockade, we focus on the statistic properties of clockwise mode photons, which are described by the zero-delay-time second order correlation function of the steady state, defined by
\begin{equation}\label{eq9}
g_{\mathrm{C}}^{(2)}(0)=\frac{\langle \hat{a}_{\mathrm{C}}^\dag\hat{a}_{\mathrm{C}}^\dag \hat{a}_{\mathrm{C}} \hat{a}_{\mathrm{C}}\rangle}{\langle \hat{a}_{\mathrm{C}}^\dag \hat{a}_{\mathrm{C}}\rangle^2}=\frac{\mathrm{Tr}\left(\rho_{ss}\hat{a}^\dag_{\mathrm{C}}\hat{a}^\dag_{\mathrm{C}}\hat{a}_{\mathrm{C}}\hat{a}_{\mathrm{C}}\right)}{[\mathrm{Tr}(\rho_{ss}\hat{a}^\dag_{\mathrm{C}}\hat{a}_{\mathrm{C}})]^2}.
\end{equation}
This physical quantity emphasizes the joint probability of detecting two photons at the same time. The value of $g_{\mathrm{C}}^{(2)}(0)<1$ ($g_{\mathrm{C}}^{(2)}(0)>1$) corresponds to sub-Poisson (super-Poisson) statistics of the cavity field, which is a nonclassical (classical) effect. This effect of the sub-Poisson photon statistics is often referred to as photon antibunching. 

\subsection{\label{sec:level2}Weak driving limit}
If the driving field coupling $F$ is very weak, due to photon blockade, only lower energy levels of the cavity field are occupied (the total excitation number of the system doesn't exceed 2). In this case, the truncated state of the system can be expanded as
\begin{align}\label{eq10}
\nonumber
|\psi\rangle=&C_{00}|00\rangle+C_{10}|10\rangle+C_{01}|01\rangle \\
&+C_{11}|11\rangle+C_{20}|20\rangle+C_{02}|02\rangle.
\end{align}
Here $|mn\rangle$ represents the fock state basis of the system with the number $m$ denoting the photon number in clockwise cavity mode, $n$ denoting the photon number in anti-clockwise cavity mode. $C_{mn}$ stands for the probability amplitude and $|C_{mn}|^2$ denotes occupying probability in the state $|mn\rangle$.
Using Eq.(\ref{eq9}) and Eq.(\ref{eq10}), the second order correlation function $g_{\mathrm{C}}^{(2)}(0)$ can be approximately given as
\begin{equation}\label{eq11}
g_{\mathrm{C}}^{(2)}(0) \simeq \frac{2|C_{20}|^2}{|C_{10}|^4}.
\end{equation}
The result of Eq.(\ref{eq11}) can be used to approximately describe the photon statistical properties in the weak driving limit. To obtain the coefficients $C_{mn}$ in Eq.(\ref{eq10}), we can substitute the state $|\psi\rangle$ into the Schro\"{o}dinger's equation $i\frac{\partial}{\partial t}|\psi\rangle=\widetilde{H}|\psi\rangle$, where $\widetilde{H}=\hat{H}_{\mathrm{sys}}-i\frac{\gamma_{\mathrm{opt}}}{2}(\hat{a}_{\mathrm{C}}^\dag \hat{a}_{\mathrm{C}}+\hat{a}_{\mathrm{A}}^\dag \hat{a}_{\mathrm{A}})$.
Then, we get a set of equations for the coefficients 
\begin{subequations}
\begin{align}
&i\frac{\partial}{\partial t}C_{00}\simeq 0,\\  
&i\frac{\partial}{\partial t}C_{10}=iF C_{00}+\bar{\Delta}C_{10}+J_2C_{01}-i\sqrt{2}F C_{20},\\ 
&i\frac{\partial}{\partial t}C_{01}=J_1C_{10}+\bar{\Delta}C_{01}-iF C_{11},\\ 
&i\frac{\partial}{\partial t}C_{11}=iFC_{01}+2\bar{\Delta}C_{11}+\sqrt{2}J_1C_{20}+\sqrt{2}J_2C_{02},\\ 
&i\frac{\partial}{\partial t}C_{20}=i\sqrt{2}FC_{10}+\sqrt{2}J_2C_{11}+2(\bar{\Delta}+U_1)C_{20},\\ 
&i\frac{\partial}{\partial t}C_{02}=\sqrt{2}J_1C_{11}+2(\bar{\Delta}+U_2)C_{02},
\end{align}
\end{subequations}
where $\bar{\Delta}=\Delta-i\frac{\gamma_{\mathrm{opt}}}{2}$.

Under the weak driving condition $F\ll\gamma_{\mathrm{in}}$, we have $|C_{00}|\gg|C_{10}|, |C_{01}|\gg|C_{20}|, |C_{11}|, |C_{02}|$, thus $C_{20}$ and $C_{11}$ can be removed in the Eq.(14b) and Eq.(14c). The vacuum state $C_{00}$ approximately has unity occupancy. Then, the steady-state solution can be found by solving the coupled equations for the coefficients. For simplicity of presentation, only $C_{10}$ and $C_{20}$ are given as below:
\begin{align}\label{eq13}
\nonumber
&C_{10}=\frac{-iF\bar{\Delta}}{\bar{\Delta}^2-J_1J_2}, \\ 
&C_{20}=\frac{1}{2\sqrt{2}}\frac{F^2[J_1J_2U+2\bar{\Delta}^2(\bar{\Delta}+U)]}{(\bar{\Delta}^2-J_1J_2)[J_1J_2(\bar{\Delta}+U)-\bar{\Delta}(\bar{\Delta}+U)^2]}.
\end{align}
With Eq.(\ref{eq11}) and Eq.(\ref{eq13}), we can approximately obtain the analytical expression of the second order correlation function and the optimal condition for the photon blockade. However, the exact expressions for the condition $g_{\mathrm{C}}^{(2)}(0)\approx 0$ are too cumbersome to be presented here. Interestingly, from Eq.(\ref{eq13}), it is obvious that the second order correlation function is closely related to the relative angular position of two particles. In particular, in the vicinity of exceptional points, i.e., $\beta=\beta_c\approx0.4$, $g_{\mathrm{C}}^{(2)}(0)\simeq |\bar{\Delta}|^2/|\bar{\Delta}+U|^2$. It means that the system shows stronger photon antibunching effect as the Kerr nonlinearity increases. When $\beta\neq\beta_c$, the case becomes different. The in-depth discussions and results of numerical calculation by the master equation approach for different parameter conditions are presented in the following subsections.
\begin{figure}
\begin{center}
\includegraphics[width=0.45\textwidth]{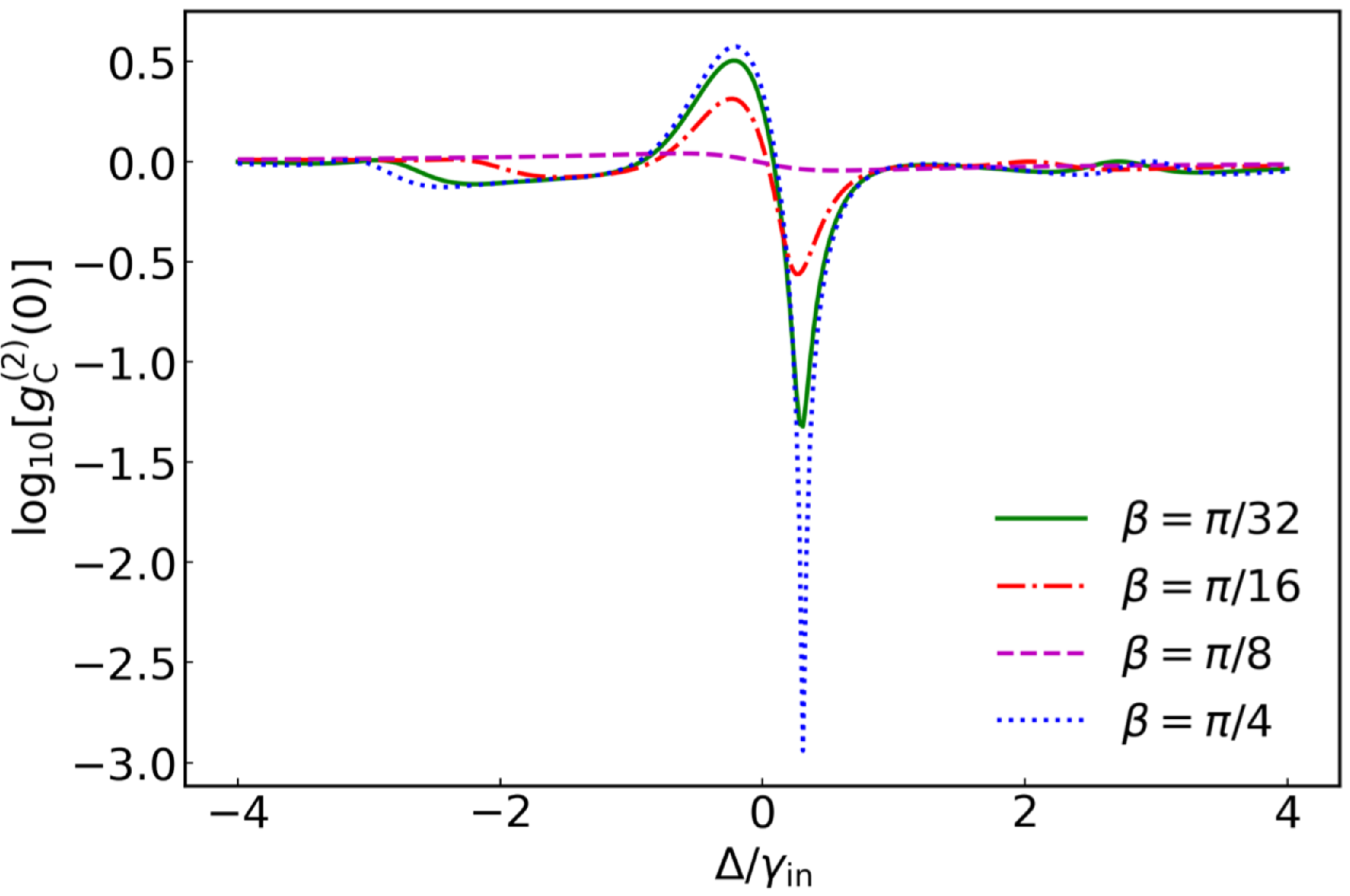}
\end{center}
\caption{The second order correlation function $\mathrm{log}_{10}[g_{\mathrm{C}}^{(2)}(0)]$ as a function of the detuning $\Delta/\gamma_{\mathrm{in}}$ under various relative angular positions $\beta$ of two nanoparticles. The parameters have been selected the same as in Fig.\ref{fig2}.}
\label{fig3}
\end{figure}

\subsection{\label{sec:level2}Single photon blockade}
In this subsection, we study the photon statistical properties of the nonlinear WGM microresonator system with two nanoparticles by numerically solving the master equation (\ref{eq7}). Figure 3 displays the second order correlation function $g_{\mathrm{C}}^{(2)}(0)$ of the cavity mode $\hat{a}_{\mathrm{C}}$ in a logarithmic scale as a function of the detuning $\Delta/\gamma_{\mathrm{in}}$ under various relative angular positions $\beta$ of two nanoparticles. Here, we consider the weak Kerr nonlinearity $U/\gamma_{\mathrm{in}}=0.059$. We can see that the profile of the second-order correlation function in a logarithmic scale $\mathrm{log}_{10}[g_{\mathrm{C}}^{(2)}(0)]$ varying with the detuning $\Delta/\gamma_{\mathrm{in}}$ exhibits a peak-dip structure. With the increasing of the detuning $\Delta$, the value of $\mathrm{log}_{10}[g_{\mathrm{C}}^{(2)}(0)]$ first arrives at the maximum and then at the minimum. For $\beta=\pi/4$, the maximum value at the peak is about $0.5$ while the minimum value at the dip is about $-3.0$. Interestingly, the photon statistics can be changed dramatically by tuning the value of relative angular position $\beta$. We find that the value of the second order correlation function $g_{\mathrm{C}}^{(2)}(0)$ is about $0.92$ when $\Delta/\gamma_{\mathrm{in}}=0.3$ and $\beta=\pi/8$. However, by tuning the parameter $\beta$ to $\pi/4$ and keeping $\Delta/\gamma_{\mathrm{in}}=0.3$, the value of $g_{\mathrm{C}}^{(2)}(0)$ rapidly decreases to  $0.002$, which indicates the strong antibunching effect. 
\begin{figure}
\begin{center}
\includegraphics[width=0.46\textwidth]{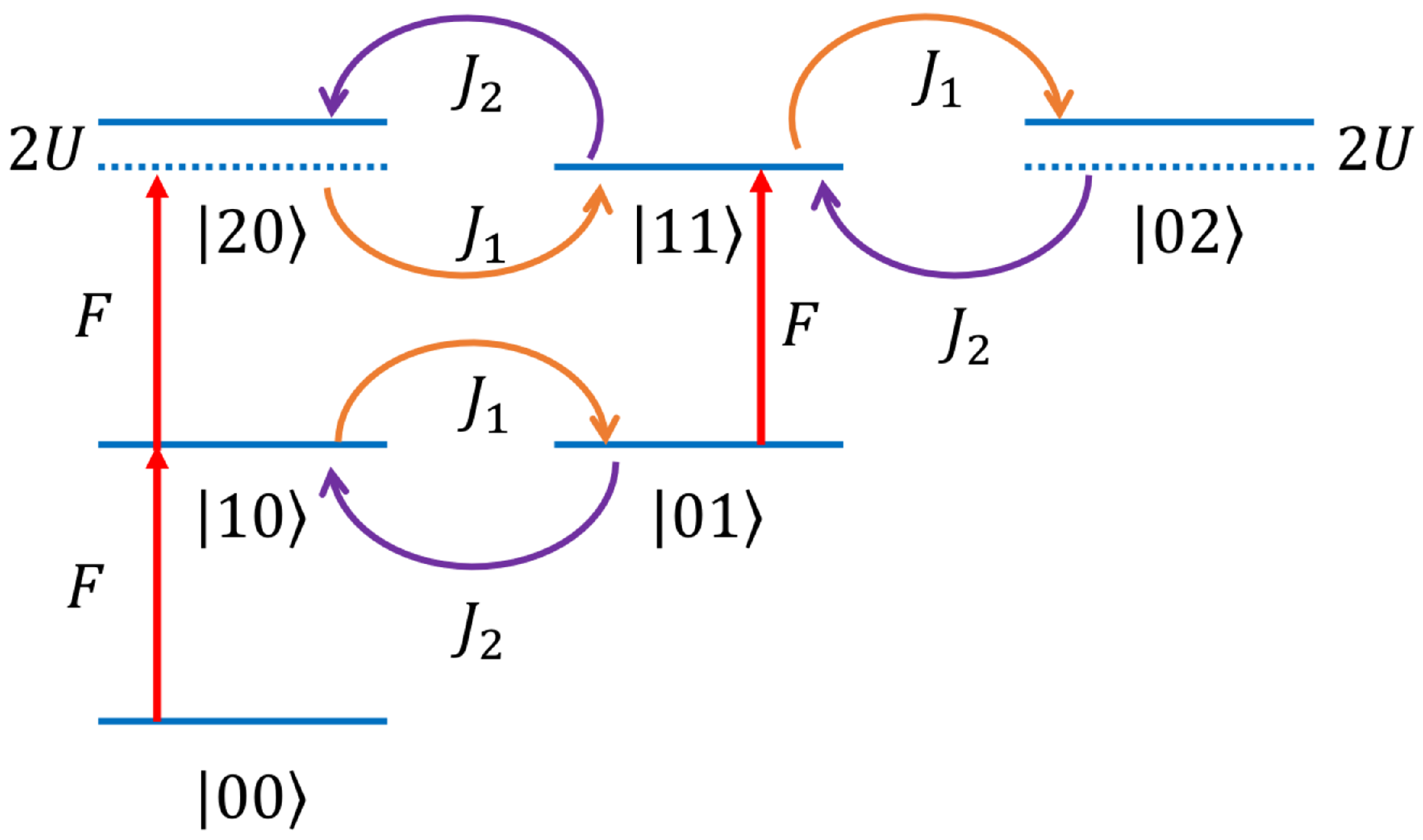}
\end{center}
\caption{Energy-level and transition path diagram of the WGM system with two nanoparticles. The quantum interference between different paths leads to strong antibunching effect.}
\label{fig4}
\end{figure}

The physical grounds behind the photon antibunching under the weak Kerr effect can be explained by the effect of quantum interference between different pathways, as shown in Fig.\ref{fig4}. There are two paths for the system to reach the two photon state of clockwise cavity mode :(i) the direct path, i.e., $|00\rangle\stackrel{F}{\longrightarrow}|10\rangle\stackrel{F}{\longrightarrow}|20\rangle$, and (ii) tunnel-coupling-mediated transition $|00\rangle\stackrel{F}{\longrightarrow}|10\rangle\stackrel{J_1}{\longrightarrow}|01\rangle\stackrel{F}{\longrightarrow}|11\rangle\stackrel{J_2}{\longrightarrow}|20\rangle$. With proper choice of parameters, the photons coming from the two pathways would destructively interfere. In other words, the destructive quantum interference between the direct path and the indirect path can reduce the probability in the two-photon excited state, this is known as unconventional photon blockade. In present model, strong antibunching effect can be achieved through adjusting the relative phase of coupling coefficients $J_1$ and $J_2$, instead of increasing the amplitudes of  them. Thus, the relative phase $\beta$ plays a crucial role in the photon statistical properties of the system. In particular, when the system is steered close to an exceptional point (i.e., $\beta\approx0.4$), the indirect path $|00\rangle\stackrel{F}{\longrightarrow}|10\rangle\stackrel{J_1}{\longrightarrow}|01\rangle\stackrel{F}{\longrightarrow}|11\rangle\stackrel{J_2}{\longrightarrow}|20\rangle$ is blocked due to the fact that $J_1=0$ or $J_2=0$ in the vicinity of the exceptional point. Only the direct path to the two photon state is allowed and then strong antibunching requires large nonlinearities, which is just the feature of the conventional photon blockade. Accordingly, a controllable switching between the unconventional and conventional photon blockade can be realized by tuning the relative angle $\beta$. 

\begin{figure}
\begin{center}
\includegraphics[width=0.42\textwidth]{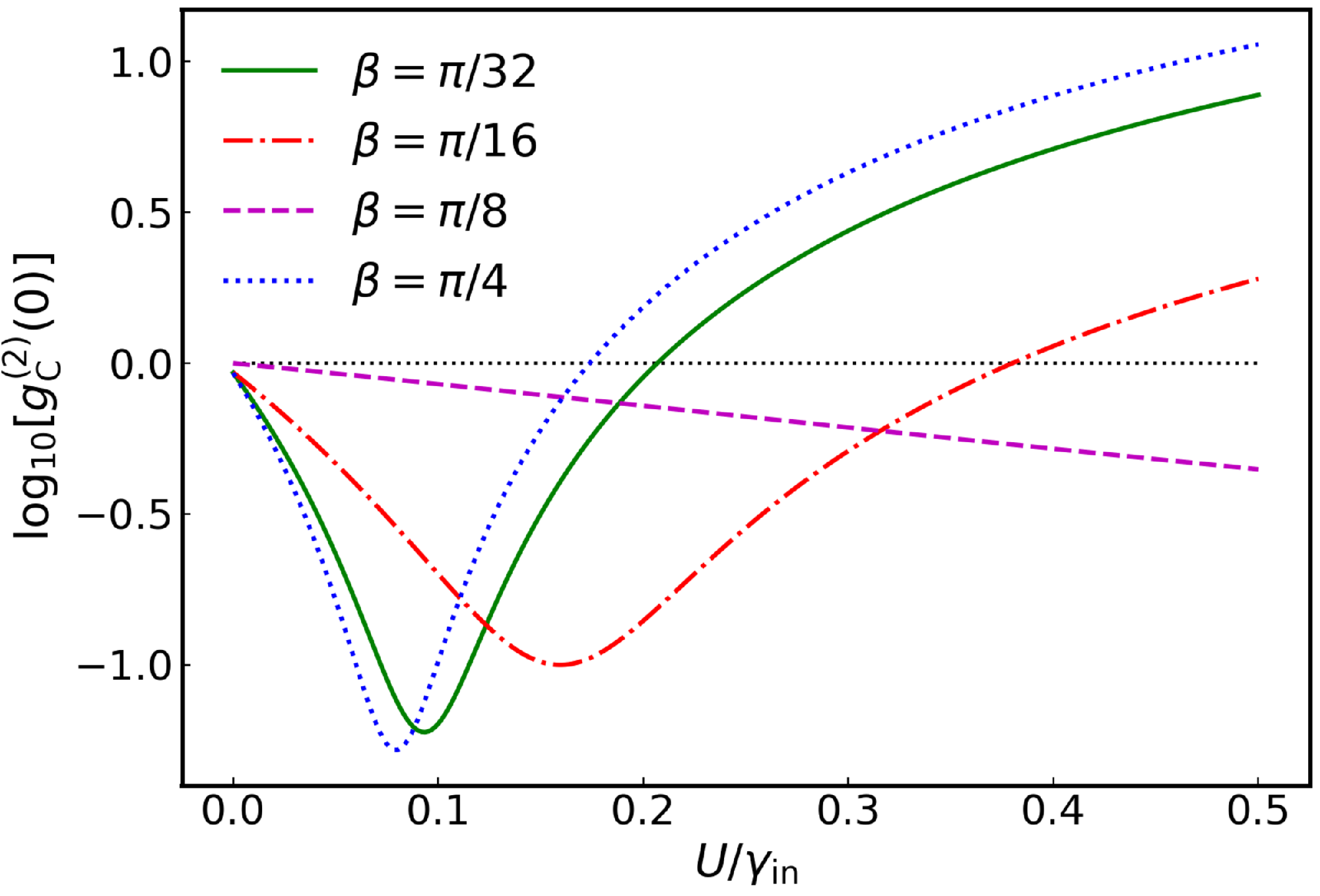}
\end{center}
\caption{The second order correlation function in a logarithmic scale $\mathrm{log}_{10}[g_{\mathrm{C}}^{(2)}(0)]$ versus Kerr nonlinearity $U/\gamma_{\mathrm{in}}$ under various values of $\beta$ by fixing $\Delta/\gamma_{\mathrm{in}}=0.4$. All other parameters are given the same as in Fig.\ref{fig2}. The black dotted line denotes the position where $\mathrm{log}_{10}[g_{\mathrm{C}}^{(2)}(0)]=0$.}
\label{fig5}
\end{figure}

In our model, another factor affecting the photon statistical properties is the Kerr nonlinearity. Figure \ref{fig5} plots the second order correlation function $g_{\mathrm{C}}^{(2)}(0)$ in a logarithmic scale versus Kerr nonlinearity $U/\gamma_{\mathrm{in}}$ under various values of $\beta$ by fixing the value of detuning at $\Delta/\gamma_{\mathrm{in}}=0.4$. In contrast to the conventional photon blockade, the value of $g_{\mathrm{C}}^{(2)}(0)$ does not always monotonically decrease with the increase of the strength of Kerr nonlinearity. It's worth noting that there exists a local minimum value of $g_{\mathrm{C}}^{(2)}(0)$, which can be adjusted by tuning the value of $\beta$. It suggests that the photon antibunching can be further enhanced with an optimal choice for the relative position $\beta$ and Kerr coefficient $U$. In previous works~\cite{25,26,27,30}, achieving strong photon antibunching with weak Kerr effect requires a large coupling strength between cavity modes. Here, we only need to tune the relative angular position $\beta$ of two nanoparticles without requiring the large coupling strength $J_{1,2}$.  Note that, in the vicinity of the exceptional points, i.e. $\beta=\pi/8$, the local minimum in the curve disappears. With increasing the strength of Kerr effect, the value of $g_{\mathrm{C}}^{(2)}(0)$ monotonically decreases. The physical reason is that, at the exceptional point, quantum interference between different pathways is  suppressed and the second order correlation function $g_{\mathrm{C}}^{(2)}(0)$ shows the features of conventional photon blockade.

\begin{figure}
\begin{center}
\includegraphics[width=0.49\textwidth]{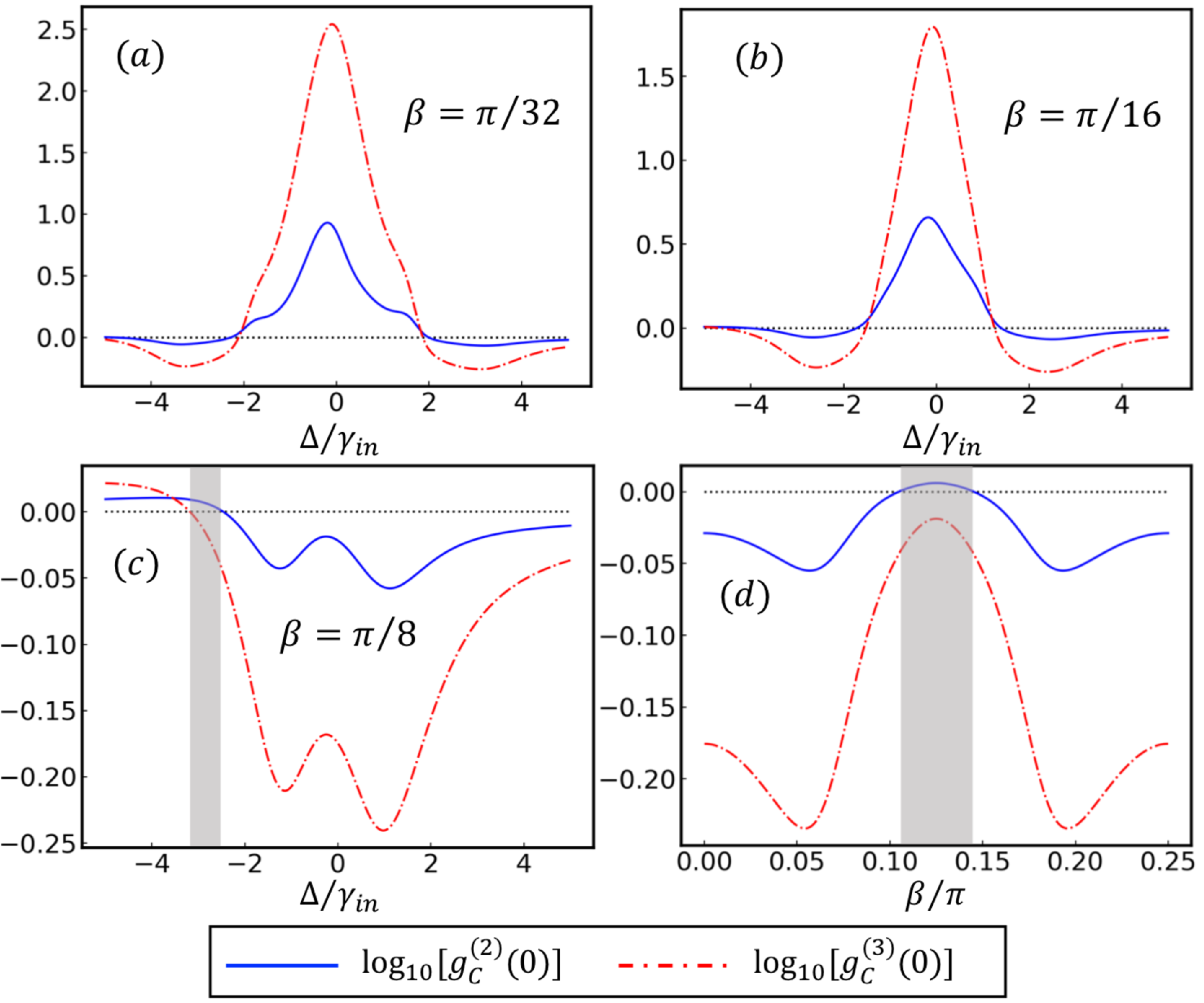}
\end{center}
\caption{The second order (blue solid curve) and third order (red dot-dashed curve) correlation function as a function of the detuning $\Delta/\gamma_{\mathrm{in}}$ under various relative positions $\beta$ in (a)-(c); (d) The second and third order correlation function as a function of relative positions $\beta$ by fixing the detuning $\Delta/\gamma_{\mathrm{in}}=-2.8$. Here we have selected $F/\gamma_{\mathrm{in}}=2$. All other parameters are given the same as in Fig.\ref{fig2}. The black dotted line denotes the position where $\mathrm{log}_{10}[g_{\mathrm{C}}^{(n)}(0)]=0$ ($n=2,3$). The two-photon bunching and three-photon antibunching can be achieved in the grey area.}
\label{fig6}
\end{figure}
\subsection{\label{sec:level2}Two photon blockade}
\begin{figure}
\begin{center}
\includegraphics[width=0.4\textwidth]{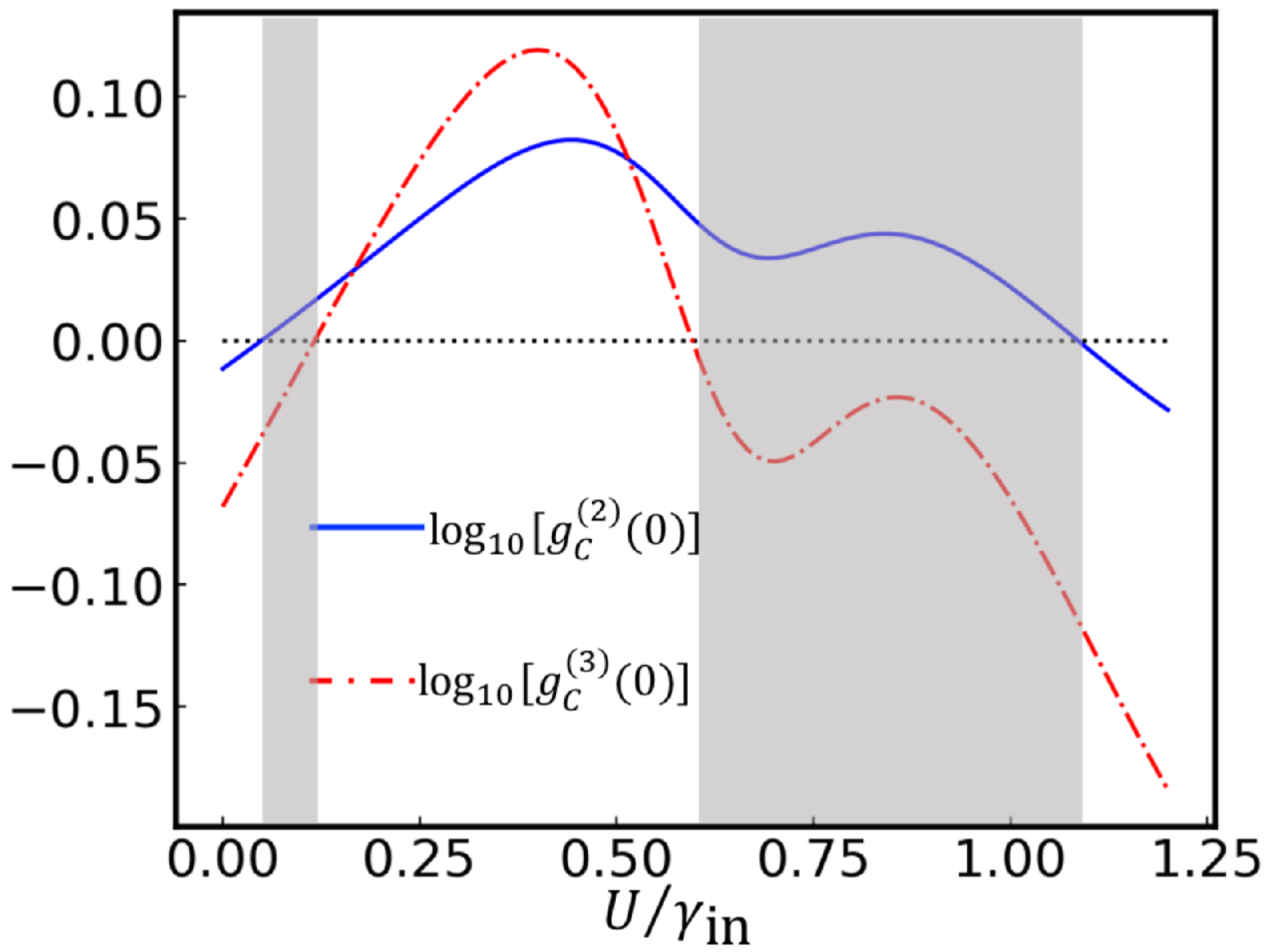}
\end{center}
\caption{The second order (blue solid curve) and third order (red dot-dashed curve) correlation function as a function of the Kerr nonlinearity strength $U/\gamma_{\mathrm{in}}$ by setting $\Delta/\gamma_{\mathrm{in}}=-2.8$, $\beta=\pi/8$, $F/\gamma_{\mathrm{in}}=2$. All other system parameters used here are the same as in Fig.\ref{fig2}. The black dotted line denotes the position where $\mathrm{log}_{10}[g_{\mathrm{C}}^{(n)}(0)]=0$ ($n=2,3$). The two-photon blockade can be achieved in the grey area.}
\label{fig7}
\end{figure}

Next, we consider the strong-pumping case (e.g., $F=2\gamma_{\mathrm{in}}$), where the photon excitation becomes much stronger. This allows for the implementation of two-photon blockade where the presence of two photons suppresses the addition of further photons. To demonstrate the two photon blockade effect, we plot the equal-time second order field correlation function ($g_{\mathrm{C}}^{(2)}(0)=\langle \hat{a}_{\mathrm{C}}^{\dag 2}\hat{a}_{\mathrm{C}}^2\rangle/\langle \hat{a}_{\mathrm{C}}^\dag\hat{a}_{\mathrm{C}}\rangle^2$) and third order field correlation function ($g_{\mathrm{C}}^{(3)}(0)=\langle \hat{a}_{\mathrm{C}}^{\dag 3}\hat{a}_{\mathrm{C}}^3\rangle/\langle \hat{a}_{\mathrm{C}}^\dag\hat{a}_{\mathrm{C}}\rangle^3$) in logarithmic units as a function of the normalized detuning $\Delta/\gamma_{\mathrm{in}}$ under various values of $\beta$ in Fig.\ref{fig6}(a)-(c). Here, the system parameters are chosen as the same as those used in Fig.\ref{fig2}. It is noteworthy that, in the vicinity of the exceptional points (i.e. $\beta=\pi/8\approx \beta_c$), clear signatures of two photon blockade phenomena ($g_{\mathrm{C}}^{(2)}(0)>1$, and $g_{\mathrm{C}}^{(3)}(0)<1$) are shown in the grey area of Figure.\ref{fig6}(c). When $\beta\neq\beta_c$, the two photon blockade phenomena disappear and at the mean time single photon blockade appears. 
The physical reason is that when the system is not near the exceptional points destructive quantum interference between different pathways leads to strong photon antibunching, so the two photon bunching is greatly suppressed. On the contrary, at the exceptional points, the two-photon bunching and three-photon antibunching can be realized under the strong driving because of the uneven energy levels of the system. 
This feature leads to an optical switching from the single-photon blockade to the two-photon blockade by just tuning the relative angular position of two nanoparticles. To show this switching operation, we plot the second order (blue solid curve) and third order  (red dot-dashed curve) field correlation functions as a function of the relative angular position $\beta/\pi$ in Fig.\ref{fig6}(d) by fixing the detuning $\Delta/\gamma_{\mathrm{in}}=-2.8$. Moreover, Kerr effect are also crucial for the degree of three photon antibunching. Figure \ref{fig7} plots the second order and third order field correlation functions versus Kerr nonlinearity strength by fixing the relative angular position $\beta=\pi/8$ and detuning $\Delta/\gamma_{\mathrm{in}}=-2.8$. From fig.\ref{fig7}, we find that two photon blockade effect occurs in the grey region. Therefore, the choice of relative position $\beta$ and Kerr nonlinearity $U$ is very important for achieving the two photon blockade in the system discussed here.

\section{conclusions}
In conclusion, we have studied the photon statistical properties in the nonlinear WGM microresonator coupled with two nanoparticles. By tuning the relative angular position $\beta$ of two nanoparticles, the photon statistical properties of the system can be well controlled and the switching between unconventional and conventional photon blockade can be achieved. We also investigate the influence of the Kerr effect on the second order correlation function and find that there is an optimal choice for relative position $\beta$ and Kerr coefficient $U$ to generate strong antibunching. Moreover, under the strong driving, two photon bunching and three photon antibunching can be achieved when the system is steered to the exceptional points. Compared with previous schemes (i.e. requiring large optical coupling between cavity modes), our scheme presented here is a good  candidate for the realization of nonclassical light generation using small nonlinearities and weak driving fields. 

\begin{acknowledgments}
We acknowledge the anonymous referees for their constructive comments to improve the paper. This work was supported by the National Natural Science Foundation of China (NSFC) (No. 61604045 and 11604059), the Natural Science Foundation of Guangdong Province, China (2017A030313020), and the Scientific Research Project of Guangzhou Municipal Colleges and Universities (1201630455).
\end{acknowledgments}

\end{document}